# ON THE CHALLENGES OF ANGULAR MOMENTUM TRANSFER MECHANISM OF PULSAR GLITCHES

**I. O. Eya**[1,3,*], **C. I. Eze**[2,3], **E. U. Iyida**[2,3]

[1]Department of Science Laboratory Technology, University of Nigeria, Nsukka
[2]Department of Physics and Astronomy, University of Nigeria, Nsukka
[3]Astronomy and Astrophysics Research Lab, Faculty of Physical Sciences University of Nigeria, Nsukka
*Corresponding Author: innocent.eya@unn.edu.ng

**Abstract**

The angular momentum transfer mechanism of pulsar glitches for over four decades now stands as standalone theory with regards to glitch sizes and inter-glitch time intervals. However, recent analyses both analytic and statistical are on the verge of suppressing the theory. In this paper, we overviewed the superfluid theory to highlight the efficacy of the angular momentum process and highlighted the theoretical and observational evidences that could not be explained by processes involving transfer of angular momentum.

## 1. Introduction

Compact astronomical objects such as neutron stars provide specimens where matter exists in extraordinary conditions not found here on earth. Understanding the interior of neutron star is of utmost importance to astrophysicists as it presents a situation where matter exist in ultra-nuclear densities, close to that in the early universe. In interior of neutron star; matter can exhibit superfluidity and superconductivity at temperature far above room temperature ($\approx 10^{10} K$). In such extreme physical condition, exotic particles such as quarks, pions and kaons exist in free state (Pandharipande and Smith, 1975; Weber et al., 2007). Generation of gravitational wave is also possible in such situations (Sedrakian et al., 2008; Cornish and Littenberg, 2015). These interesting facts could only be reproduced in our local universe if scientists understand the actual nature of interior of a neutron star. Astrophysicists are interested in pulsar glitches, as they offer one the opportunity to probe the interior of neutron star. Pulsar glitches refer to the sudden spin-up of pulsars without the effects of external force, there-by linking glitches with neutron star interior processes and behaviour of matter at exotic states. Understanding of the mechanisms involved in pulsar glitches shall pave way for nuclear physicists in exploring matter at super-nuclear densities. Unfortunately, pulsar glitches are rare phenomenon, coupled with the poor understanding of superfluidity and superconductivity at extreme temperatures comparable to that in interiors of neutron stars. These effects are some of the hindrances in understanding the actual nature of neutron star interior. Nonetheless, theorists have used pulsar glitches to constrain some physical properties of neutron star such as mass (Ho et al., 2015), and radius (Link et al., 1999; Steiner et al., 2015; Eya et al., 2016b).

The origins of glitches in pulsars are still under serious investigation despite numerous models to explain them (see Haskell and Melatos (2015) for a recent review). After the first pulsar glitch, many models were propounded to explain the origin, but for the fact that the pulse signal remained invariant during and after the event, models involving magnetospheric activities (Michel, 1970; ScargleandPacini, 1971) were abandoned. But now, variation in pulse profile and other magnetospheric activities now accompany some glitch events especially those involving magnetars and few other high magnetic field pulsar (Dib et al., 2008; Livingstone et al., 2011; Weltevrede et al., 2011; Keith et al., 2013). Presently, models used in interpreting pulsar glitches are broadly classified into two based on the mechanism driving them. They are: the starquake and the angular momentum transfer models.

Starquake model relies on sudden reduction in neutron stars' moment of inertia (Ruderman, 1969: 1991; Baym and Pines, 1971). In this model, it is assumed that a fast spinning neutron star is more oblate than spherical, and that the surface of the star is a crystalline solid (crust) which can support stress. As the star spins down, stress builds up in the crust. At a certain critical stress, starquake occurs resulting in sudden reduction in the stellar moment of inertia, leaving the star with a less oblate shape. The sudden reduction in moment of inertia results in sudden spin-up of the crust known as glitch. This model is convenient in explaining small size glitches (e.g. Crab pulsar glitches) but fails to explain large glitches (e.g. Vela pulsar glitches) (Flanagan, 1996).

In the angular momentum transfer model, the assumption is that neutron stars consist of at least two components in relative motion (Baym et al., 1969). The components are the solid crust and the interior superfluid. The latter is regarded as angular momentum reservoir. A spinning superfluid is believed to be threaded with an array of vortices. The number of vortices it contains per unit area is proportional to its angular velocity. In the case of pulsars, these vortices are believed to be pinned to the inner layer of the crust (Anderson and Itoh, 1975; Alpar, 1977). As long as these vortices remain pinned (or their number density constant), the superfluid component is notexpected to spin-down. However, the crust spins-down due to electromagnetic braking torque acting on it. This process results in differential rotation lag with the crust lagging behind. As the lag grows with time, at a certain critical lag, unknown mechanism unpins the vortices or alters their position; the unpinned vortices migrate outward and transfer their momentum to the crust. When this happens, the superfluid spins-down and the crust spins-up — glitch (Anderson and Itoh, 1975; Alpar et al., 1984). A glitch happening in this way should follow a well defined order in the sense that accumulation and release of momentum is a function of pulsar rotational parameters only. This scenario has been observed in PSRs, $J0835 − 4510$, and $J0537 − 6910$ (McCulloch et al., 1987; Link et al., 1999; Middleditch et al., 2006; EyasndUrama, 2014; Eya et al., 2017).

After a glitch, a recovery phase usually follows during, which the pulsar regain a steady spin frequency. In some cases, incomplete recovery leading to a new spin frequency and frequency derivative has been observed (Lyne, 1987; Lyne et al., 1992, 1993). The recovery phase is linear in some pulsars and mostly exponential in others (Yu et al., 2013). Exponential recovery as a process is a good indicator that the motion of neutron stars mimics the motion of a container and a fluid content, instead of that of a rigid body. The recovery time is too long (weeks to months) when compared to the time predicted by viscous forces (hours). These observations are strong indications that a component of zero viscosity preferable superfluid neutrons is involved in the glitch process (Lyne and Graham-Smith, 1998), there-by supporting angular momentum transfer process as a mechanism of pulsar glitch. In addition to the idea of superfluid component supporting the angular momentum transfer mechanism, the agreement of the theoretical magnitude of neutron star crustal fluid (momentum reservoir) moment of inertia (Ravenhall and Pethick 1994) with that predicted from glitch data (Link et al., 1999) made the angular momentum transfer mechanism a standard for analysing pulsar glitches. However, in recent works, this age long standard is under serious challenge with the involvement of neutrons entrainment on the mobility of superfluid (Chamel, 2012). Physically, neutron entrainment encodes the mobility of superfluid neutrons. In this, Andersson et al., (2012) involved entrainment on the magnitude of superfluid neutrons participating in glitch and declared that the angular momentum reserved in the crustal fluid is not enough to produce the observed glitch size. In contrast to this, other works show that the *crust can be enough* if one explores the uncertainties in equation of states to model a neutron star with a larger crustal thickness (Piekarewicz et al., 2014; Steiner et al., 2015). Larger crust can accommodate more fluid, which can produce large glitches even with the entrainment factor.

Meanwhile, since the interpretation of pulsar glitches as a process of exchange of momentum between neutron star components (Baym et al., 1969; Anderson and Itoh, 1975), efforts are on going to know the location of superfluid component participating in glitch. The location of superfluid component participating in glitch depends on the structure of the neutron star which depends on its size (i.e. radius (R)) and mass. Unfortunately, direct measurement of any of these quantities is virtually impossible with present day instruments. Theorists have been using Equation of State (EoS) to estimate both quantities (e.g. Baym et al. 1971; Pandharipande and Smith, 1975; Friedman and Pandharipande, 1981). A major problem with this approach is that the radius of the star (likewise the thickness of the crust) depends on how you want the EoS to be. In this frame, the magnitude of the superfluid component depends on the EoS used. This questions the veracity to a certain extent of the process as one gets different magnitudes for different EoSs.

In this brief review, we present an overview of the superfluid theory to give credence to the existence of bulk rotation of superfluid via quantized vortices and examine the mechanism involved in the angular momentum transfer. Finally, both the theoretical and observational evidences challenging the angular momentum mechanism are presented.

## 2. Overview of Superfluid Theory

Unlike classical fluids, a superfluid is a coherent fluid of zero entropy and viscosity (Alexander and Jeerold, 1997); one of the realization of a Bose-Einstein condensate. This enables its dynamics to be explained by a single particle wave function $\psi$ as in London model called the condensate wave function (London, 1938a,b). Apart from the neighbourhood of a rigid container or in a vortex core, the superfluid density is a slow varying function of position. As a result, a bulk of the fluid may be regarded as being of uniform density. In addition, superfluid neutrons are paired in a manner analogous to the formation of cooper pairs in a superconductor. The paired neutrons then form a boson. This result in fluid having *bosonic* properties that leads to formation of Bose-Einstein condensate. London described the condensate by a means of wave function of the form:

$$\psi(\vec{r},t) = \psi_0(\vec{r},t)\exp(is(\vec{r},t)) \qquad (1)$$

This he assumed to be a solution of the Schroedinger's wave equation (Landau & Lifshitz 1987). To obtain the superfluid velocity $\vec{V}_s$, one applies the momentum operator, which is of the form:

$$i\hbar \nabla \psi = \vec{P}\psi, \qquad (2)$$

but

$$\vec{P} = m_{np}\vec{V}_s, \qquad (3)$$

where $m_{np}$ is the mass of the neutron pair. From Equations (2) and (3), one obtains the superfluid velocity

$$\vec{V}_s = (\hbar/m_{np})\nabla S, \qquad (4)$$

where S is the phase, which is a function of position. For conservation of mass, the continuity equation of the form:

$$\frac{\partial \rho_s}{\partial t} = -\nabla \cdot \vec{j}_s, \qquad (5)$$

is introduced, where $\rho_s$ is the superfluid density and $\vec{j}_s$ superfluid current density. For the population density, one has

$$\psi^2(\vec{r},t) = \frac{\rho_s}{m_{np}} \qquad (6)$$

Furthermore, Troup (1967) showed that for large particle number N, one can state:

$$\partial N \partial S \approx 1 \qquad (7)$$

meaning that N and S can be approximated to conjugate variable, therefore the equations of motion of this quantity are

$$\frac{\hbar \partial S}{\partial t} = \frac{-\partial H}{\partial N}, \qquad (8)$$

and

$$\frac{\hbar \partial N}{\partial t} = \frac{-\partial H}{\partial S} \qquad (9)$$

where H is the Hamiltonian, which is defined as

$$H = H_{s,k} + H_0 \qquad (10)$$

$H_{s,k}$ is the kinetic energy of the superfluid and $H_0$ is the potential energy. Using mean values, the equation of motion for S becomes

$$\hbar \frac{\hbar \partial S}{\partial t} = \frac{-\partial H}{\partial N} = -\left(\mu + \frac{1}{2}m_{np}V_s^2\right) \qquad (11)$$

where $\mu$ is the chemical potential defined as

$$\mu = \left(\frac{\partial H_0}{\partial N}\right)_{entropy, volume} \tag{12}$$

Then using the gradient of Equation (11) and with Equation (4) one arrives at superfluid equation of motion, thus

$$\hbar \frac{\partial \nabla S}{\partial t} = m_{np} \frac{\partial \vec{V}}{\partial t} = -\nabla\left(\mu + \frac{1}{2} m_{np} V_s^2\right), \tag{13}$$

exploring the convective derivative defined as

$$\frac{D\vec{a}}{Dt} = \frac{\partial \vec{a}}{\partial t} + (\vec{v} \cdot \nabla)\vec{a} \tag{14}$$

for an arbitrary vector $\vec{a}$ in a constant volume element travelling in a velocity field $\vec{v}$ the Euler equation for an ideal fluid (a fluid of just zero viscosity) is given as

$$\frac{D\vec{V}_s}{Dt} = \frac{\partial \vec{V}_s}{\partial t} + (\vec{V}_s \cdot \nabla)\vec{V}_s = -\frac{\nabla \mu}{m_{np}} \tag{15}$$

Also, exploring the properties of $\nabla$ operator gives

$$\frac{D\vec{V}_s}{Dt} = \frac{\partial \vec{V}_s}{\partial t} + \nabla(\frac{1}{2}V_s^2) - \vec{V}_s \times (\nabla \times \vec{V}_s) = -\frac{\nabla \mu}{m_{np}} \tag{16}$$

Comparing Equation (13) and (16) gives,

$$\nabla \times \vec{V}_s = 0 \tag{17}$$

if

$$\vec{V}_s \neq 0 \tag{18}$$

Equation (17) is the Landau criterion for superfluidity. This equation has a serious consequence in rotation of superfluid. Meanwhile, the circulation of superfluid is defined as

$$k = \oint_l \vec{V}_s \cdot dl \tag{19}$$

for an integration over a contour $l$ that is wholly in superfluid and bearing in mind of the Landau criterion and using Stokes' theorem in Equation (19), one obtains

$$k = (\nabla \times \vec{V}_s)dA = 0. \tag{20}$$

The end result of Equation (20) is that the rotation of pure superfluid is impossible as a consequence of Landau criterion. This result (i.e. Equation (20)) has been confirmed experimentally by Andronikashvili (1946). In that experiment, oscillating pile of disks entrained the normal fluid component while leaving the superfluid component at rest. But the mechanism of angular momentum transfer in pulsar glitch is hinged on rotation of superfluid neutrons. Why? A critical examination of Equation (19) shows that it is possible for the circulation of superfluid, and hence its rotation depending on the homogeneity of the medium. If the region inside the contour of the integration is to be multiple connected, that is, having a vortex core – a cylindrical region inside bulk superfluid devoid of fluid or

35

containing normal fluid or free neutrons where $\nabla \times \vec{V_s} \neq 0$, the circulation is possible. Using Equation (4) and (19), leads to expression for circulation of superfluid in terms of phase.

$$k = \frac{\hbar}{m_{np}} \oint_l \nabla S . dl = \frac{\hbar}{m_{np}} (\nabla S)_l \qquad (21)$$

From Equation (1), which is single valued, the value of such a wave motion around a contour should be invariant. Therefore, due to the nature of the wave function, the possible changes in the values 'S' are multiples of $2\pi$ and zero. This implies that values of zero corresponds to the Landau criterion, while that of $2\pi$ gives non-zero circulation, and hence apply to the case of vortices. In this case, the circulation is given by

$$k = n \frac{\hbar}{m_{np}}, \qquad (22)$$

where n is an integer. Accordingly, the circulation is quantised. For that reason, superfluid rotation can occur in macroscopic realm in the presence of multiple connected regions formed by quantized vortex lines. If we consider a streamline at radius r from the center of an isolated vortex line, the following equation applies

$$k = \oint_l \vec{V_s} . dl = 2\pi V_s(r) \qquad (23)$$

Combining Equation (22) and (23) gives the velocity of the fluid at a given distance from the vortex core as

$$\vec{V_s} = \frac{k}{2\pi r} = \frac{n\hbar}{m_{np}} \qquad (24)$$

then the angular momentum at that point is

$$L = m_{np} V_s r = n\hbar \qquad (25)$$

Equation (25) means that the angular momentum is also quantized. Therefore, in addition to macroscopic circulation, both the velocity and angular momentum of the fluid around a vortex line are quantised.

## 3. Angular Momentum Transfer Mechanism

This mechanism involves angular momentum transfer between two differentially rotating neutron star components. The components involved are the interior superfluid and the solid crust as well as all other portion of the star strongly coupled to the crust. Superfluid rotates by formation of array of vortices, which carry the circulation. In neutron star, these vortices are pinned in the inner-crust (Anderson and Itoh, 1975) which leads to partial decoupling of the bulk fluid from the star. In this picture, the superfluid velocity is constant and higher than that of the solid crust as it is not affected by electromagnetic braking torque on the crust. The process enables the superfluid to store angular momentum. The standard differential rotation lag is given as

$$\omega(t) = v_s - v_c \qquad (26)$$

while the conservation of stellar total moment of inertia is

$$I = I_c + I_s \qquad (27)$$

where $v_s$ is the superfluid spin frequency, $v_c$ is the crust (pulsar) spin frequency, $I_c$ and $I_s$ are the crustal and interior superfluid moment of inertia respectively. A change in angular momentum of the crust ($2\pi\Delta I_c v_c$) during glitch resulted from a corresponding change in angular momentum of the superfluid ($2\pi\Delta I_s v_s$). So at the onset of the glitch, angular momentum conservation entails

$$I_c(v_c + \Delta v_c) = I_s(v_s - \Delta v_s) \tag{28}$$

Please note that the change in spin frequency of the superfluid is in negative direction as the superfluid is required to spin-down for the crust to spin-up. From equation (28), one obtains

$$I_c v_c - I_s v_s = -I_s \Delta v_s - I_c \Delta v_c \tag{29}$$

For complete transfer of momentum in one direction (i.e. $\Delta v_c = -\Delta v_s$), the glitch size is

$$\Delta v_c = \frac{I_s}{I} v_s - v_c \tag{30}$$

Also note that wehave explored the expression of the total moment of inertia, and used $I_c \approx I$ as the stellar core is strongly coupled to crust (Link et al., 1999, and references therein). This process can give rise to all glitch sizes inasmuch as the superfluid velocity ($v_s$) is larger than that of the crust ($v_c$) as $I_s/I$ can at most be equal to unity

## 4. Challenges Associated with Mechanism of Angular Momentum
### 4.1 Neutron star crustal thickness and neutron entrainment

It is believed that the bulk of superfluid is ir-rotational but in multiple connected regions as seen in Equation (21) the rotation is feasible. The rotation/flow rate is not a function of time or does it depend on the texture of the container. Instead, it has to do with the number of vortex core (or vortices) available [(Equation (22)], as such its macroscopic flow with zero viscosity is understandable. Likewise, the superfluid neutrons contained in the inner-crust of a neutron superfluid rotate with zero viscosity. Though the viscosity is zero, the mobility is still entrained in the crust (Chamel and Carter, 2006; Pethick et al., 2010; Chamel, 2012). Entrainment is non-dissipative in nature; it is a result of elastic scattering of free neutrons[†] by the crustal lattice (Chamel, 2013). With the density of free neutrons (Chamel, 2013) or its effective mass (Andersson et al. 2012), the magnitude of the entrainment is readily quantified. The magnitude of entrainment viewed as entrainment factors limits the observed glitch size especially the large ones with respect to the magnitude of neutron star components participating in glitches. Using the average glitch activity and mean spin-down rate in a pulsar, to introduce the pulsar characteristic age in other to get the accumulated spin-down rate of the crust (Andersson et al., 2012) the glitch size is

$$\frac{\Delta v}{v} \approx \frac{m_n}{m_n^*}\left(\frac{I_n}{I}\right)\frac{t_{glitch}}{2\tau_c}, \tag{31}$$

which leads to

$$\frac{I_n}{I} \approx 2\tau_c A \frac{\langle m_n^* \rangle}{m_n}, \tag{32}$$

---

[†] These free neutrons are also one of the reasons while the integral over the contour [Equation (19)] could be multiple connected.

in expressing the entrainment in terms of the (average) effective neutron mass, $\langle m_n^* \rangle$, and the bare neutron mass, $m_n$, (Andersson et al., 2012). The implication of this is that in the limit where, $\langle m_n^* \rangle \gg m_n$, the entrainment lowers the effective moment of inertia associated with the superfluid and that the magnitude of the superfluid in the inner-crust is not enough to account for large pulsar glitches as the size is constrained by Equation (31).

Moreover, more recent analyses show that Fractional Moment of Inertia (FMI) of neutron star components participating in glitch is (Eya et al., 2017)

$$\frac{I_s}{I_c} = -\frac{1}{\dot{v}(t)} \frac{\Delta v_c(t)}{t_i}, \quad (33)$$

And the total angular momentum of spinning neutron star as a result of neutron entrainment in the inner crust is (Eya et al., 2019)

$$L_{tot} = I_{ss}(v_s - v_c) + I v_c \quad (34)$$

where $I_{ss}$ is the entrained neutrons moment of inertia. The term in the parenthesis is the differential rotation lag $\omega(t)$, [see Equation(26)], implying that the moment of inertia associated with transferred momentum is that of the entrained neutron. It follows that on incorporating entrainment factor on the individual glitch FMI (Eya et al. 2019)

$$\frac{I_s}{I_c} = -\frac{1}{E_n \dot{v}_c(t)} \frac{\Delta v_c(t)}{t_i}, \quad (35)$$

Equation (35) indicates that the observed glitch sizes need to be lowered by a factor of $\frac{1}{E_n}$, in order to be accommodated in the momentum reservoir container, or that the size of the container $I_s$ needs to be increased by a factor of $E_n$. An analysis of the distribution of FMIs of glitch sizes using Equation (33) indicates that with or without neutron entrainment, the magnitude of neutron star crustal superfluid is not enough to produce large glitches (Basu et al., 2018; Eya et al., 2019). This effect becomes severe when Equation (35) is used (Eya et al., 2019), as no present day EoS can model such a neutron star that could contain the required superfluid for the larger glitches.

### 4.2 Radiative changes accompanying glitches

Formerly, it is due to lack of radiative changes associated with pulsar glitches that gave credence to internal origin of pulsar glitches and subsequently to the angular momentum transfer mechanism. However, evidences of radiative changes accompanying glitch events have been reported in magnetars (Dib et al., 2008; Dib and Kaspi, 2014; Kaspi and Beloborodov, 2017) and recently in radio pulsars (Keith et al., 2013; Kou et al., 2018). In magnetars, the radiative changes appear as X–ray bursts from the pulsar magnetosphere. These bursts could not be explained by magnetar model. Magnetars are just young pulsars powered by the decay of their ultra high magnetic field unlike radio pulsars powered by the loss of their rotational kinetic energy. With respect to radio pulsars of similar characteristic age, glitches in magnetars should be smaller than what is observed, as such, contribution of the burst to their glitch sizes could be feasible (Gao et al., 2016). If this is proved observationally, angular momentum transfer mechanism will not properly account for the glitch size.

On the other hand, in radio pulsars, glitches accompanied with radiative changes inform of change in pulse profile and spin-down state have also been observed in PSR J1119-6127 (Weltevrede et al., 2011), PSRs J0742-2822 and J2021+4026 (Keith et al., 2013; Zhao et al., 2017). In addition to the change in spin-down state, emission mode changes and narrowing of pulse associated with glitch activity was also observed in PSR B2035+36 (Kou et al., 2018). In this pulsar, the change in spin-down state occurred as a persistent increase in $|\dot{v}|$ for over 800 days after the glitch coupled with emission mode switching. This post glitch behaviour is not in line with glitch recovery process, which is normally interpreted as a restoration of equilibrium between the faster spinning interior superfluid and the crust. As such the angular momentum transfer mechanism could not interpret all aspect of the glitch.

### 4.3 Anti-glitches in Magnetar

Anti-glitches are generally seen in timing data as sudden spin-down in pulsar spin frequency. Although these events have been observed earlier in radio pulsars (Cordes and Downs, 1985; Cordes et al., 1988; Chukwude and Urama, 2010); it only attracted attention of researchers when it started occurring in magnetars (Archibald et al., 2013) with features, which could not be ordinarily seen as consequences of sudden spin-down. In any model of neutron star, it is believed that the interior of neutron star contains superfluid, which is weakly coupled to other components. As such, the interior superfluid should be rotating faster than the outer component affected by electromagnetic braking torque. Any sudden coupling of the components will lead to spin-up of the crust known as glitch. But for the crust to suddenly spin-down, it will require the crust to suddenly couple to a component, which is actually spinning slowly compared to the crust. This component does not exist anywhere in the neutron star. In addition, hard X-ray bursts also accompany the anti-glitches, which could not be explained with the standard glitch model. The only way employed to explain this observation is the collision of magnetar with small body, which has angular momentum that is anti-parallel to the magnetar (Huang and Geng, 2014). Though this process could readily account for the X–ray burst, due to the evaporation of the small body after collision, it could not well explain the glitch recovery process which mimics that of angular momentum transfer process.

### 5. Discussion

The angular momentum transfer mechanism of pulsar glitches is under serious challenge due to observational and theoretical evidences, which is not in en tandem with the process. The earlier agreement between theory and glitch data (Link et al., 1999) that gave credence to the angular momentum transfer mechanism has been questioned seriously with recent theoretical findings. The question now is whether pulsars are neutron stars, and if they are, the present day EoS is not sufficient to model a glitching pulsar. Conventionally, lack of radiative change associated with early glitches in rotation-powered pulsars is one of the strongest indicators that causes of pulsar glitches have more to do with the internal mechanism of pulsars. However, noticeable changes in pulse profile linked with timing irregularity in a number of pulsars are pointers that their rotations are interrelated with their emissions (Lyne et al., 2010). However, for now, it is not yet well understood what induced the radiative changes, though glitch may be a trigger mechanism. Kou et al. (2018) in explaining the possible cause of magnetospheric activity observed in PSR B2035+36 glitch, suggests that it could be due to change in external braking torque resulting from the glitch event. A change in external braking torque could result from uneven out-flowing particle density in the magnetosphere (Kou and Tong, 2015), which could lead to the persistent increase in spin-down rate. Another suggestion is that glitch may cause a change in the inclination angle[‡] that could alter the magnetic field structure (Ng et al., 2016), thereby leading to a persistent increase in the spin-down rate. Kou et al. (2018) also pointed out that sudden fluctuation of the inclination angle shall manifest in the change of effective emission geometry, hence the observed pulse profile variation. Then if this processes, which are still theoretical could be triggered by glitch; another school of thought is that these processes could also trigger glitch. In that, glitches associated with radiative changes, which could not be explained with the standard glitch model is understandable. In conclusion, the angular momentum transfer mechanism is no longer enough to explain glitches following recent findings and challenges associated with it. It is therefore necessary for astrophysics to search/develop other models to account for the challenges.

---

[‡] the angle between magnetic axis and rotational axis